\documentclass[12pt]{article}

\usepackage{amsfonts,amsmath,amssymb}
\usepackage{bm}
\usepackage{cite}

\allowdisplaybreaks[3]

\newcommand{\R}{\mathbb{R}}
\newcommand{\Z}{\mathbb{Z}}
\newcommand{\Q}{\mathbb{Q}}

\newcommand{\sfrac}[2]{{#1/#2}}

\newcommand{\correctedA}[1]{{#1}} 
\newcommand{\correctedB}[1]{{#1}} 

\begin{document}

\begin{titlepage}

\begin{flushright}
KUNS-2304\\
YITP-10-89
\end{flushright}

\vspace{0.1cm}

\begin{center}
  {\Large
  Emergent AdS$_3$ in the Zero Entropy Extremal Black Holes
  }
\end{center}

\vspace{0.1cm}

\begin{center}
         Tatsuo {\sc Azeyanagi}$^{\spadesuit}$\footnote
           {
E-mail address : aze@gauge.scphys.kyoto-u.ac.jp}, 
         Noriaki {\sc Ogawa}$^{\heartsuit}$\footnote
         {
E-mail address : noriaki@yukawa.kyoto-u.ac.jp} and    
         Seiji {\sc Terashima}$^{\heartsuit}$\footnote
           {
E-mail address : terasima@yukawa.kyoto-u.ac.jp}

\setcounter{footnote}{0}

\vspace{0.3cm}

${}^{\spadesuit}$
{\it Department of Physics, Kyoto University,\\
Kyoto 606-8502, Japan}\\

${}^{\heartsuit}$
{\it Yukawa Institute for Theoretical Physics, Kyoto University,\\
     Kyoto 606-8502, Japan}
\end{center}

\vspace{1.5cm}

\begin{abstract}
We investigate the zero entropy limit of the
near horizon geometries of $D=4$ and $D=5$ general extremal black holes 
with $\mathrm{SL}(2,\R)\times \mathrm{U}(1)^{D-3}$ symmetry.
We \correctedA{derive some conditions on the geometries from expectation of regularity.}
We \correctedA{then} show that \correctedA{an} AdS$_3$ structure emerges
in \correctedA{a certain scaling} limit,
though the periodicity shrinks to zero.
\correctedA{We present} some examples to see \correctedA{the above} concretely.
We also comment on some implications to the Kerr/CFT correspondence.
\end{abstract}
\end{titlepage}

\tableofcontents

\section{Introduction}
\label{introduction}

The (generalized) Kerr/CFT correspondence
\cite{Guica:2008mu,Lu:2008jk,Azeyanagi:2008kb,%
Hartman:2008pb,Compere:2009dp,Azeyanagi:2009wf}
is one of the promising candidates for the framework to
understand quantum aspects of black holes
in a broader class, including realistic ones.
It has already provided a fairly general framework for finding the dual theories
(which are 2D chiral CFT's) to extremal black holes,%
\footnote{
Generalizations 
of the Kerr/CFT to non extremal black holes are also proposed
and the existence of dual non-chiral CFT's is suggested%
\cite{Bredberg:2009pv,Cvetic:2009jn,Castro:2010fd}.
However, appropriate boundary conditions have not been found for it yet,
although there are some attempts in this direction%
\correctedA{\cite{Matsuo:2009sj,Matsuo:2009pg,Rasmussen:2009ix,Rasmussen:2010sa}}.
}
not limited to the ones embedded
in string or supersymmetric theories.
Deeper understanding and wider generalizations of the Kerr/CFT would
reveal some general aspects and practical applications
of gauge/gravity correspondence and quantum gravity.

However, the understanding of the Kerr/CFT is still rather poor
and there are many seemingly strange assumptions in it.
One of the most important mysteries of the Kerr/CFT is 
why it works well itself.
Unlike the AdS/CFT's based on brane setup in string/M theories\cite{Maldacena:1997re},
the interpretation of the dual chiral CFT is, generically, quite obscure.
Furthermore, the boundary condition imposed on the geometry is rather violent,
which implies that the dual CFT does not correspond to one fixed
macroscopic background, but a series of many backgrounds with
macroscopically different charges\cite{Azeyanagi:2008dk}.
It in turn suggests that there is no dynamics in a fixed extremal background,
as is pointed out in both holographic%
\cite{Azeyanagi:2008dk,Balasubramanian:2009bg}
and gravitational%
\cite{Amsel:2009ev,Dias:2009ex}
points of view.
This fact also makes difficult a clear understanding of the Kerr/CFT.

Against the problem above, a hopeful conjecture proposed up to now is
that the Kerr/CFT comes from AdS$_3$/CFT$_2$ correspondence%
\cite{Azeyanagi:2008dk,Balasubramanian:2009bg}.
It has been argued that some AdS$_3$ structures would be hidden
behind the near horizon geometries of extremal black holes,
and that the chiral CFT$_2$ would appear as some limit of the non-chiral
CFT$_2$ dual to the AdS$_3$\cite{Strominger:1997eq},
though no explicit realizations were proposed.

Very recently, a related and very interesting result was reported\cite{Guica:2010ej}.
They investigated the maximal charge limit
of 5D extremal Kerr-Newman (or BMPV\cite{Breckenridge:1996is,Breckenridge:1996sn,Cvetic:1996xz}) black hole%
\cite{Guica:2007gm,Gibbons:1999uv,AlonsoAlberca:2002wr,Dyson:2006ia},
where the entropy goes to zero,
and 
showed that the total space including graviphoton fiber
in the near horizon limit
is locally the same as the one for the BPS black string at zero left and right
temperatures, AdS$_3\times \mathrm{S}^3$.
They then argued that the Kerr/CFT (with central charge $c=6J_L$)
for this system
is embedded in string theory which provides the microscopic realization
\cite{Strominger:1996sh,Maldacena:1997de}
under the maximal charge limit,
and also discussed some deformation from there.%
\footnote{
  For a recent related attempt, see \cite{Compere:2010uk}.
}

{}From this, on one hand,
one may hope that there is an AdS$_3$ structure in some points 
in the parameter space of general extremal black holes
with rotational symmetries,
although their simple realizations in string theory are difficult
to imagine generally.
On the other hand,
the 5D black holes considered in \cite{Guica:2010ej}
can be uplifted to the 6D black string solution with an AdS$_3$
structure in the near horizon geometry,
which is not expected for general extremal black holes.
Thus, the emergence of AdS$_3$ is seemingly due to the special 
properties of the solutions.

In this paper, however, 
we find that the AdS$_3$ structure indeed emerges 
just by taking the zero entropy limit, $S_{\mathit{BH}}=0$,
of the near horizon geometries
of almost general \correctedA{extremal} black holes with axial $\mathit{U}(1)^{D-3}$
in $D=4$ and $D=5$.
Here, the zero entropy limit, we mean, does not include
the small black holes and massless limits
\correctedA{--- that is,}
the
\correctedA{near horizon $\mathrm{AdS}_2$ structure must not collapse.}

\correctedA{
{}From a holographic point of view,
an extremal black hole should correspond to the ground states of
a superselection sector \correctedA{with fixed charges} in the boundary theory.
Therefore, some decoupled infrared theory is expected to live there,
even when the degeneracy of the ground states vanishes.
It suggests the existence of some scaling limit
where the near horizon geometry remains regular%
\footnote{\correctedA{
Strictly speaking, this ``regular'' means ``regular almost everywhere''.
It will always be the case henceforth in this paper.
}}
while the entropy goes to zero.
This expectation requires some additional conditions on
the general form of the near horizon geometry\cite{Kunduri:2007vf,Hollands:2009ng,Kunduri:2010vg} in this zero entropy limit.
In fact, these conditions are satisfied in the concrete examples
which we will investigate later.%
}

\correctedA{Those conditions, in turn, prove to lead to}
the emergence of an AdS$_3$ structure.%
\footnote{\correctedA{
This emergence of AdS$_3$ is the same one as has already been observed
in some special systems, in \cite{Guica:2010ej,Guica:2007gm,Gibbons:1999uv,AlonsoAlberca:2002wr,Dyson:2006ia} above and also in \cite{Nakayama:2008kg}.
In each of them, the near horizon limit is taken first,
and after that the zero entropy limit is taken.
On the other hand, a similar but inverse-ordered limit is investigated
for some black holes
in \cite{Bardeen:1999px} and \cite{Balasubramanian:2007bs,Fareghbal:2008ar}, for example.
In those cases, an AdS$_3$ structure does also emerge,
but in a slightly different manner.
}}
\correctedA{We will see it in the concrete examples, too, and finally}
discuss some implications of our theorem to the Kerr/CFT.

Organization of this paper is as follows.
In \S\ref{4d_general} and \S\ref{5d_general}, we start with general form of 
near horizon geometry of extremal black holes 
with $\mathrm{SL}(2,\R)\times \mathrm{U}(1)^{D-3}$ isometry 
for $D=4$ and $D=5$ cases respectively.
We then derive
\correctedA{
some conditions on the geometry in the zero entropy limit
from the expectation of regularity,%
}
and show
\correctedA{that they lead to}
the emergence of an AdS$_3$ structure\correctedA{.}
In \S\ref{Examples}, we consider the zero entropy limit 
for some concrete examples, including
5D Myers-Perry black hole\correctedA{\cite{Myers:1986un}},
5D Kaluza-Klein black hole\correctedA{\cite{Gibbons:1985ac,Rasheed:1995zv,Larsen:1999pp}}
and extremal black holes in 5D supergravity.
In \S\ref{Implications_to_KerrCFT},
following the argument of \cite{Guica:2010ej},
we explain some relation between chiral CFT$_2$ appearing in 
the Kerr/CFT and non-chiral CFT$_2$ expected to be dual 
to AdS$_3$ emerging in the zero entropy limit.         
In \S\ref{Conclusions_and_Discussions},
we end up with the conclusions and discussions.

\section{Four Dimensions}
\label{4d_general}

First we investigate the 4D extremal black holes.
In \S\ref{4d_general_zeroentropy_limit}
we will derive
\correctedA{some conditions on the behavior of the near horizon geometry in}
the zero entropy limit%
\correctedA{, based on the expectation that the geometry should remain regular.}
In \S\ref{4d_general_AdS3_emergence}
we will show that
\correctedA{those conditions guarantee that}
an AdS$_3$ structure always emerges as a covering space in the limit.

\correctedA{
Note again that, we do not consider the cases of massless or
small black holes, 
where the near horizon $\mathrm{AdS}_2$ structure collapses.
For example, Kerr and Kerr-Newman black holes are excluded,%
\footnote{\correctedA{
Kerr black hole can, though, be embedded in 5D as
``Kerr black string'',
that is, Kerr $\times\mathrm{S}^1$.
It can be regarded as a fast rotating Kaluza-Klein
black hole with the Kaluza-Klein electric and magnetic charges $Q=P=0$.
By increasing $Q$ and $P$ from there, we can obtain a zero entropy
black hole.}}
since they become inevitably massless
when the entropy goes to zero.
We will give one of the simplest examples in 4D
at the beginning of \S\ref{Examples}.
}

\subsection{Near horizon geometry}

We consider the near horizon geometry of
the extremal and axial symmetric black holes.
We assume the near horizon metric has the form as follows:%
\footnote{
We turned the sign of $k$ from \cite{Kunduri:2007vf,Kunduri:2008rs,Kunduri:2008tk},
so that it agrees with that of the rotation in terms of $\phi$.
}$^,$%
\footnote{\correctedB{
The functions $A(\theta)$, $B(\theta)$, $F(\theta)$ and the constant $k$ are not arbitrary, because
we assume that the geometry satisfies the equation of motion 
with appropriate matters. 
}}
\begin{align}
  ds^2=g_{\mu\nu}dx^{\mu}dx^{\nu}
  &=A(\theta)^2
   \left[
     -r^2dt^2 + \frac{dr^2}{r^2}
     + B(\theta)^2(d\phi - krdt)^2
   \right]
   + \correctedA{F(\theta)^2}d\theta^2,
   \label{4d_ansatz}
\end{align}
where $A(\theta), B(\theta), F(\theta) > 0$,
$0\le\theta\le\pi$ and $\phi\sim\phi+2\pi$.
This geometry has an AdS$_2$ structure with an $\mathrm{S}^1$ fiber on it,
and in turn it fibrates on an interval parameterized by $\theta$.
This ``standard form'' of the near horizon metric
is proven to be the general one
for arbitrary theories with Abelian gauge fields and uncharged scalars,
including higher-derivative interactions in general\cite{Kunduri:2007vf},
so it is quite a general one.
\correctedB{We will focus on the case without higher-derivative interactions
for simplicity.}
The Bekenstein-Hawking entropy $S_{\mathit{BH}}$ for \eqref{4d_ansatz}
is given as
\begin{align}
  S_{\mathit{BH}}
  = \frac{\mathrm{Area}(\mathrm{horizon})}{4G_4}
  = \frac{\pi}{2G_4}\int\!\!d\theta\, A(\theta)B(\theta)F(\theta),
  \label{4d_entropy}
\end{align}
where $G_4$ is the 4D Newton constant.
On the other hand, the volume element $d^4V$ is
\begin{align}
  d^4V = \sqrt{-g}\,d^4x = A(\theta)^3B(\theta)F(\theta)dtdrd\theta d\phi.
  \label{4d_volume}
\end{align}

\subsection{Regularity conditions in zero entropy limit}
\label{4d_general_zeroentropy_limit}

Now we consider the zero entropy limit, $S_{\mathit{BH}}\to 0$,
for the geometry \eqref{4d_ansatz}.
We are interested in black holes, and so the geometry
should remain both regular and nontrivial in this limit.
For this purpose, rescalings of the coordinates are allowed in general.
But along the angular directions,
we look at the whole region of the geometry since we regard it
as a black hole.
Therefore we focus on the scale where, generically,
$d\theta\sim 1,\,d\phi\sim 1$.%
\footnote{
In this paper, $X\sim Y$ means that $\lim\sfrac{X}{Y}$ is
a nonzero finite value, while
$\lim\sfrac{X}{Y}$ may be $0$ for $X = \mathcal{O}(Y)$.
Although we also use $\sim$ to describe periodic identifications,
they can usually be distinguished clearly from the context.
}
Furthermore, since \eqref{4d_ansatz} has a scaling symmetry
(as a subgroup of $\mathrm{SL}(2,\R)$)
\begin{align}
  t\to \frac{t}{\lambda},\quad r\to \lambda r,
\end{align}
for an arbitrary constant $\lambda$, we can fix the scale of
the coordinate $t$ by this transformation.
So we always take $dt\sim 1$.
Under these conditions above for scales,
$A(\theta)\sim 1,\, F(\theta)\sim 1$
is obviously required
to prevent the geometry from collapse.
Therefore we have to take $B(\theta)\to 0$ for $S_{\mathit{BH}}=0$.
But at the same time, the volume element \eqref{4d_volume}
has to remain nonzero and finite.
The only way under the current conditions is to take
\begin{align}
  B(\theta) = \epsilon B'(\theta),
  \quad r = \frac{r'}{\epsilon},
  \quad B'(\theta)\sim 1,
  \quad r'\sim 1,
  \quad \epsilon\to 0.
\end{align}
By using these new variables,
\eqref{4d_ansatz} becomes
\begin{align}
  ds^2
  &= A(\theta)^2
   \left[\frac{dr'^2}{r'^2}
    + \Big(B'(\theta)^2k^2-\frac{1}{\epsilon^2}\Big)r'^2dt^2
    - 2B'(\theta)^2\epsilon kr'dtd\phi
    + B'(\theta)^2\epsilon^2d\phi^2\right]
   + F(\theta)^2d\theta^2 \nonumber\\
  &= A(\theta)^2
   \left[\frac{dr'^2}{r'^2}
    + \Big(B'(\theta)^2k^2-\frac{1}{\epsilon^2}\Big)r'^2dt^2
    - 2B'(\theta)^2\epsilon kr'dtd\phi\right]
   + F(\theta)^2d\theta^2 + \mathcal{O}(\epsilon^2).
  \label{4d_regular_metric_1}
\end{align}
In order that the geometry does not collapse,
the $dtd\phi$ term has to remain nonzero and it implies
\begin{align}
  k = \frac{k'}{\epsilon},
  \quad k'\sim 1.
  \label{4d_kprime}
\end{align}
However, in that case the $dt^2$ term becomes
\begin{align}
  \Big(B'(\theta)^2k^2-\frac{1}{\epsilon^2}\Big)r'^2dt^2
  = \epsilon^{-2}(B'(\theta)^2k'^2-1)r'^2dt^2,
\end{align}
which generically diverges.
We can escape from this divergence only when
\begin{align}
  B'(\theta)k'
  = 1 + 
    \epsilon^2b(\theta),
  \quad b(\theta)=\mathcal{O}(1),
  \label{4d_b}
\end{align}
that is, $B'(\theta)$
should go to a $\theta$-independent constant $B'\equiv\sfrac{1}{k'}$
in the $\epsilon\to 0$ limit.
In terms of the original parameters,
the zero entropy limit consistent with regularity implies
\begin{align}
  B(\theta)\to 0,\quad k\to\infty,\quad
  \text{while}\quad B(\theta)k\to 1.
\label{4d_zeroentropy_limit}
\end{align}
Using \eqref{4d_kprime} and \eqref{4d_b},
the metric \eqref{4d_regular_metric_1} is finally written
in the $\epsilon\to 0$ limit as
\begin{align}
  ds^2=A(\theta)^2
   \bigg(
   \frac{dr'^2}{r'^2}
    + 2b(\theta)r'^2dt^2
    - \frac{2}{k'}r'dtd\phi\bigg)
   + F(\theta)^2d\theta^2,
   \label{4d_regular_metric}
\end{align}
showing the regularity manifestly.
\correctedA{
(Especially, if $b(\theta)\to 0$, this metric 
 has a local AdS$_3$ structure
 in the null selfdual orbifold form\cite{Coussaert:1994tu}.)}
Then we can conclude that
\eqref{4d_zeroentropy_limit} is the general condition
for the near horizon geometry \eqref{4d_ansatz} to have vanishing
entropy while keeping itself regular.

\subsection{AdS$_3$ emergence}
\label{4d_general_AdS3_emergence}

Let us return to the original metric \eqref{4d_ansatz},
and define a new coordinate $\phi'$ as%
\footnote{
This $\phi'$ is essentially similar one to the $y$ in \cite{Guica:2010ej}.
Our $k$ corresponds to $\frac{1}{2\pi T_Q}$ there.
}
\begin{align}
  \phi' = \frac{\phi}{k},
\end{align}
whose periodicity is given by
\begin{align}
  \phi' \sim \phi' + \frac{2\pi}{k}.
\end{align}
Using this coordinate, the metric \eqref{4d_ansatz} is written as
\begin{align}
  ds^2&=A(\theta)^2
   \left[
     -r^2dt^2 + \frac{dr^2}{r^2}
     + B(\theta)^2k^2(d\phi' - rdt)^2
   \right]
   + \correctedA{F(\theta)^2}d\theta^2.
\end{align}
Under the limit \eqref{4d_zeroentropy_limit},
while formally regarding $dr\sim 1$ and $d\phi'\sim 1$,
the metric becomes
\begin{align}
  ds^2&=A(\theta)^2
   \left[
     -r^2dt^2 + \frac{dr^2}{r^2}
     + (d\phi' - rdt)^2
   \right]
   + \correctedA{F(\theta)^2}d\theta^2,
   \label{4d_AdS3}
\end{align}
where we find an (not warped or squashed) AdS$_3$
structure, fibrated on the $\theta$ direction.

Note that, however, to be really an AdS$_3$,
the coordinate $\phi'$ has to run from $-\infty$ to $\infty$,
while the period of $\phi'$ is $\sfrac{2\pi}{k}\to 0$ here.
This makes \eqref{4d_AdS3} singular,
and so the precise meaning of the form \eqref{4d_AdS3}
is quite subtle and we leave it for future investigation.
In the present stage,
we interpret our ``AdS$_3$'' above
to emerge as a covering space of the original geometry.
In other words, the AdS$_3$ is
orbifolded by an infinitesimally narrow period.
This orbifolding may be regarded as a zero temperature limit of
BTZ black hole, as was adopted in \cite{Guica:2010ej}.

\correctedA{
The AdS$_3$ structure can also be obtained directly from
the zero entropy regular geometry \eqref{4d_regular_metric}.
When we take infinitesimal $r'$,
the $dt^2$ term turns to be subleading in $r'$ expansion,
by considering the diagonalization or eigenequation for the metric.
Therefore the metric becomes
\begin{align}
  ds^2\approx A(\theta)^2
   \bigg(
   \frac{dr'^2}{r'^2}
    - \frac{2}{k'}r'dtd\phi
   \bigg)
   + F(\theta)^2d\theta^2,
   \label{4d_metric_in_rprime_leading}
\end{align}
in the first order of $r'$.
On the other hand, if we use $\phi$ again and regard $\correctedA{d}\phi\sim 1$
in \eqref{4d_AdS3},
the metric is rewritten as
\begin{align}
  ds^2&=A(\theta)^2
   \bigg(
     \frac{dr^2}{r^2}
     - \frac{2\epsilon}{k'}rdtd\phi
     + \frac{\epsilon^2}{k'^2}d\phi^2
   \bigg)
   + F(\theta)^2d\theta^2.
   \label{4d_metric_usingepsilon}
\end{align}
Similarly to the discussion above,
the $d\phi^2$ term proves to be subleading,
and so this metric also becomes 
\eqref{4d_metric_in_rprime_leading}
in the first order of $\epsilon$, by using $r'=\epsilon r$.
It means that the zero entropy regular geometry \eqref{4d_regular_metric}
itself has the infinitesimally orbifolded $\mathrm{AdS}_3$ structure
in the infinitesimal $r'$ region.}%
\footnote{\correctedB{
We can also obtain
the same form as \eqref{4d_metric_in_rprime_leading},
by defineing $t'\equiv\epsilon t$ for \eqref{4d_metric_usingepsilon}.
}}

\section{Five Dimensions}
\label{5d_general}

After the 4D case in \S\ref{4d_general},
in this section we will examine the 5D case.
Although it is more complicated than 4D case
because of the existence of two rotational directions,
we successfully show a similar theorem to 4D case,
warranting the emergence of an AdS$_3$ structure.

\subsection{Near horizon geometry}

Let us consider 5D extremal black holes with two axial symmetries.
We assume the metric has the form of \correctedB{\cite{Kunduri:2007vf}}
\begin{align}
  ds^2=g_{\mu\nu}dx^{\mu}dx^{\nu}
    &=
   A(\theta)^2
   \Big[
     -r^2dt^2 + \frac{dr^2}{r^2}
     + B(\theta)^2(d\phi_1 - k_1rdt)^2 \nonumber\\
     &\qquad
     + C(\theta)^2\big(d\phi_2 - k_2rdt + D(\theta)(d\phi_1 - k_1rdt)\big)^2
   \Big]
   + F(\theta)^2d\theta^2,
   \label{5d_ansatz}
\end{align}
where $A(\theta), B(\theta), C(\theta), F(\theta) > 0$,
$k_1, k_2\geq 0$, $0\le\theta\le\pi$
and
\begin{align}
  \phi_1\sim\phi_1 + 2\pi,
  \quad \phi_2\sim\phi_2 + 2\pi.
  \label{5d_periodicity}
\end{align}
This is proven to be the general form under the same \correctedA{condition}
as the 4D case.
The Bekenstein-Hawking entropy $S_{\mathit{BH}}$ for \eqref{5d_ansatz}
is given as
\begin{align}
  S_{\mathit{BH}}
  = \frac{\mathrm{Area}(\mathrm{horizon})}{4G_5}
  = \frac{\pi^2}{G_5}\int\!\!d\theta\, A(\theta)^2B(\theta)C(\theta)F(\theta),
  \label{5d_entropy}
\end{align}
where $G_5$ is the 5D Newton constant.
The volume element $d^5V$ is
\begin{align}
  d^5V
  = \sqrt{-g}\,d^5x 
  = A(\theta)^4B(\theta)C(\theta)F(\theta)\,dtdrd\theta d\phi_1 d\phi_2.
  \label{5d_volume}
\end{align}

\paragraph{Modular transformations}

Since $\phi_1$ and $\phi_2$ forms a torus $\mathrm{T}^2$ in the
coordinate space, we can act
the modular transformation group $\mathrm{SL}(2,\Z)$
while keeping the metric to be the form of \eqref{5d_ansatz}
and the periodicities \eqref{5d_periodicity}.
If necessary for keeping $k_1,k_2>0$.
we redefine the signs of $\phi_1$ and $\phi_2$ at the same time.
The generators $\mathcal{S}$ and $\mathcal{T}$ of this $\mathrm{SL}(2,\Z)$ are
defined by
\begin{align}
  \mathcal{S}:&\quad (\tilde{\phi}_1,\tilde{\phi}_2) = (-\phi_2,\phi_1),
  \label{5d_S_transformation:1}\\
  \mathcal{T}:&\quad (\tilde{\phi}_1,\tilde{\phi}_2) = (\phi_1,\phi_1+\phi_2),
  \label{5d_T_transformation:1}
\end{align}
respectively,
although $\mathcal{S}$ always acts together with the redefinition of the sign
of $\tilde{\phi_1}$, behaving as a mere swapping
\begin{align}
  \mathcal{S}':&\quad (\tilde{\phi}_1,\tilde{\phi}_2) = (\phi_2,\phi_1).
  \label{5d_Sp_transformation:1}
\end{align}
Under $\mathcal{S}'$ and $\mathcal{T}$,
the functions and parameters in \eqref{5d_ansatz}
are transformed as, \cite{Loran:2009cr}
\begin{align}
  \mathcal{S}':\quad
  &\tilde{B}(\theta)^2=\frac{B(\theta)^2C(\theta)^2}{B(\theta)^2+C(\theta)^2D(\theta)^2},
  \quad
  \tilde{C}(\theta)^2=B(\theta)^2+C(\theta)^2D(\theta)^2, \nonumber\\
  &\tilde{D}(\theta)=\frac{C(\theta)^2D(\theta)}{B(\theta)^2+C(\theta)^2D(\theta)^2},
  \quad
  \tilde{k}_1 = k_2,\quad \tilde{k}_2=k_1,
  \label{5d_Sp_transformation:2}\\
  \mathcal{T}:\quad
  &\tilde{B}(\theta)^2=B(\theta)^2,
  \quad
  \tilde{C}(\theta)^2=C(\theta)^2, \nonumber\\
  &\tilde{D}(\theta)=D(\theta)-1,
  \quad
  \tilde{k}_1 = k_1,\quad \tilde{k}_2=k_2+k_1.
  \label{5d_T_transformation:2}
\end{align}

\subsection{Regularity conditions in zero entropy limit}
\label{5d_general_zeroentropy_limit}

Now we consider the regularity conditions in
$S_{\mathit{BH}}\to 0$ limit for
\eqref{5d_entropy}.
Here we explain the outline and the results.
The details are given in Appendix \ref{5d_general_proof}.

{}From exactly a similar discussion to that
in \S\ref{4d_general_zeroentropy_limit},
we take
\begin{align}
  A(\theta)\sim 1,\quad F(\theta)\sim 1,
  \quad d\theta\sim 1,
  \quad d\phi_1\sim 1,
  \quad d\phi_2\sim 1,
  \quad dt\sim 1,
\end{align}
and then it is required that
\begin{align}
  B(\theta)C(\theta)\sim\epsilon,
  \quad r =\frac{r'}{\epsilon},
  \quad r'\sim 1,
  \quad \epsilon\to 0.
  \label{5d_zeroentropy_limit1}
\end{align}
Obviously, divergence of $B(\theta)$ or $C(\theta)$ causes the
$d\phi_1^2$ or $d\phi_2^2$ term of the metric \eqref{5d_ansatz}
to diverge, and so it is not allowed.
Therefore, to realize \eqref{5d_zeroentropy_limit1},
there are three possibilities for the behaviors of
$B(\theta)$ and $C(\theta)$,
depending on ether (or both) of them goes to $0$ in the limit.
However, by using the swapping transformation $\mathcal{S}'$ given in
\eqref{5d_Sp_transformation:1}\eqref{5d_Sp_transformation:2},
we can show that all the cases are resulted in the case of
\begin{align}
  B(\theta)\sim\epsilon,
  \quad C(\theta)\sim 1.
  \label{5d_BC_limit}
\end{align}
In this case, similarly to the 4D result \eqref{4d_zeroentropy_limit},
\begin{align}
  B(\theta)k_1 = 1 + \mathcal{O}(\epsilon^2)
  \label{5d_Bk=1}
\end{align}
has to be satisfied. In other words,
\begin{align}
  B(\theta)=\epsilon B'(\theta),
  \quad k_1 = \frac{k'_1}{\epsilon},
  \quad k_1'B'(\theta) = 1 + \epsilon^2b(\theta),
  \quad
  B'(\theta)\sim 1,
  \quad k_1'\sim 1,
  \quad b(\theta)=\mathcal{O}(1).
  \label{5d_Bk1_limit}
\end{align}
Finally, for remaining parameters $k_2$ and $D(\theta)$,
the condition proves to be
$D(\theta)=\mathcal{O}(1)$ and $k_2 + D(\theta)k_1=\mathcal{O}(\epsilon)$.
This is satisfied if and only if
\begin{align}
  k_2 = \frac{k'_2}{\epsilon},
  \quad D(\theta) = -\frac{k'_2}{k'_1} + \epsilon^2d(\theta),
  \quad k'_2=\mathcal{O}(1),
  \quad d(\theta)=\mathcal{O}(1).
  \label{5d_Dk2_limit}
\end{align}
Then $D(\theta)$ goes to a constant
$D\equiv -\sfrac{k'_2}{k'_1} = -\sfrac{k_2}{k_1}$,
which may or may not be $0$.

Under \eqref{5d_zeroentropy_limit1}
\eqref{5d_BC_limit} \eqref{5d_Bk1_limit} \eqref{5d_Dk2_limit},
the metric \eqref{5d_ansatz} becomes
\begin{align}
  ds^2=A(\theta)^2
   \bigg[
     & \Big(2b(\theta) + k_1'^2C(\theta)^2d(\theta)^2\Big)r'^2dt^2
     +\frac{dr'^2}{r'^2}
   \nonumber\\
     &- 2\Big(
\correctedA{
     \frac{1}{k_1'}
}
     +C(\theta)^2d(\theta)D\Big) r'dtd\phi_1
     - 2C(\theta)^2d(\theta)r'dtd\phi_2
   \nonumber\\
     &+ C(\theta)^2D^2d\phi_1^2
     + 2C(\theta)^2Dd\phi_1d\phi_2
     + C(\theta)^2d\phi_2^2
   \bigg]
   + F(\theta)^2d\theta^2,
\end{align}
which is indeed regular.
\correctedA{
It is simply rewritten by defining
\begin{align}
  \phi'_2 = \phi_2 + D\phi_1,
  \label{5d_phi2prime}
\end{align}
as
\begin{align}
  ds^2=A(\theta)^2
   \bigg[
     & \Big(2b(\theta) + k_1'^2C(\theta)^2d(\theta)^2\Big)r'^2dt^2
     +\frac{dr'^2}{r'^2}
   \nonumber\\
     &
     - \frac{2}{k_1'} r'dtd\phi_1
     - 2C(\theta)^2d(\theta)r'dtd\phi_2'
     + C(\theta)^2d\phi_2'^2
   \bigg]
   + F(\theta)^2d\theta^2.
   \label{5d_regular_metric}
\end{align}
}

\subsection{AdS$_3$ emergence}
\label{5d_general_AdS3_emergence}

Now let us look at the metric \eqref{5d_ansatz},
under the same limit
\eqref{5d_BC_limit} \eqref{5d_Bk1_limit} \eqref{5d_Dk2_limit},
for the parameters as \eqref{5d_regular_metric},
but different scalings for the coordinates.
Using the original coordinates, the metric is written as
\begin{align}
  ds^2=A(\theta)^2
   \bigg[
     -r^2dt^2
     + \frac{dr^2}{r^2}
     + \bigg(\frac{d\phi_1}{k_1} - rdt
\correctedA{
        + \frac{\epsilon^3b(\theta)}{k_1'}d\phi_1 - \epsilon^2b(\theta)rdt
}
\bigg)^2&
     \nonumber\\
     + C(\theta)^2\Big(d\phi_2
\correctedA{
     + Dd\phi_1
     + \epsilon^2d(\theta)d\phi_1- \epsilon k_1'd(\theta)rdt
}
\Big)^2&
   \bigg]
   + F(\theta)^2d\theta^2.
   \label{5d_AdS3}
\end{align}
\correctedA{
Now we switch the coordinates from $(\phi_1,\phi_2)$ to $(\phi_1',\phi_2')$,
where $\phi_2'$ is \eqref{5d_phi2prime} and $\phi_1'$ is defined as
\begin{align}
  \phi'_1 = \frac{\phi_1}{k_1}.
  \label{5d_phi1prime}
\end{align}
We}
regard $d\phi'_1\sim 1$, $d\phi'_2\sim 1$ and $dr\sim 1$,
together with $dt\sim 1$, $d\theta\sim 1$.
Then in the $\epsilon\to 0$ limit, \eqref{5d_AdS3} goes to
\begin{align}
  ds^2=A(\theta)^2
   \bigg[
     -r^2dt^2
     + \frac{dr^2}{r^2}
     + \Big(d\phi_1' - rdt\Big)^2
     + C(\theta)^2d\phi_2'^2
   \bigg]
   + F(\theta)^2d\theta^2.
   \label{5d_AdS3_manifest}
\end{align}
Manifestly, \eqref{5d_AdS3_manifest} has locally a form of a product of
$\mathrm{AdS}_3$ and $\mathrm{S}^1$, fibered on the $\theta$-interval.%
\footnote{\correctedB{
At the same time, we notice that the rotation along the $\phi'_2$
direction vanishes here and then the geometry is static.
Therefore, we can say that the near horizon geometry in
the zero entropy limit always results in the static AdS$_3$ case
classified by \cite{Kunduri:2007vf}, though 
the periodicity goes to zero here.}}

Of course, the periodicities of $\phi'_1$ and $\phi_2'$ are
problematic.
As for $\phi'_1$ in the AdS$_3$, the situation is exactly the
similar to the 4D case explained in \S\ref{4d_general_AdS3_emergence}.
What about the $\mathrm{S}^1$ coordinate $\phi'_2$\,?
When $D$ is an integer, there is no problem.
In fact, in this case we could use the modular transformation $\mathcal{T}$
\eqref{5d_T_transformation:1} \eqref{5d_T_transformation:2} (or $\mathcal{T}^{-1}$)
repeatedly, to make $D=0$ in advance.
On the other hand,
in case $D$ is not an integer, it may cause some new problem,
perhaps depending on whether $D\in\Q$ or not.
For, when we take a covering space over $\phi_1$ and have a full AdS$_3$,
the $D\phi_1$ term will do nothing on the $\mathrm{S}^1_{\phi'_2}$,
regardless of the value of $D$.

\correctedA{
In a similar way to the 4D case, the AdS$_3$ structure can be
obtained from \eqref{5d_regular_metric}.
When we take infinitesimal $r'$,
the metric becomes
\begin{align}
  ds^2=A(\theta)^2
   \bigg(
     \frac{dr'^2}{r'^2}
     - \frac{2}{k_1'} r'dtd\phi_1
     + C(\theta)^2d\phi_2'^2
   \bigg)
   + F(\theta)^2d\theta^2,
\end{align}
in the leading order of $r'$,
and this coincides with that of \eqref{5d_AdS3_manifest}.
}

\section{Examples}
\label{Examples}

In the previous \correctedA{sections}, we systematically argued
\correctedA{the form of the near horizon geometry and}
the emergence of \correctedA{the} AdS$_3$ structure
\correctedA{in the zero entropy limit}.
In this section, we consider some interesting examples of
extremal black holes
\correctedA{
to demonstrate our discussions above.
We see that the regularity conditions we derived are
indeed realized in each case,
and an AdS$_3$ emerges as a result of it.
}

\correctedA{We} deal with a class of 5D vacuum extremal black holes
and the ones in the 5D supergravity.
The former includes the extremal Myers-Perry black \correctedA{hole} and
the extremal slow rotating Kaluza-Klein black \correctedA{hole} as we will explain,
while the latter includes the setup discussed in \cite{Guica:2010ej}.

Because the examples below are complicated,
we give one of the simplest examples here.
Let us consider the 4D extremal 
slow rotating dyonic black \correctedA{hole} in Einstein-Maxwell-dilaton theory
\cite{Astefanesei:2006dd}.
In the near horizon limit, the geometry is written as
\begin{align}
  ds^2 &=
  \frac{2G_4J(u^2-1)\sin^2\theta}{\sqrt{u^2-\cos^2\theta}}
  \left(d\phi-\frac{rdt}{\sqrt{u^2-1}}\right)^2
  +2G_4J\sqrt{u^2-\cos^2\theta}
  \left(-r^2dt^2+\frac{dr^2}{r^2}+d\theta^2\right),
\end{align}
and the entropy is expressed as
$S_{BH} = 2\pi J\sqrt{u^2-1}$.
Here $\tilde{Q}$, $\tilde{P}$, $J$ are
the electric charge, the magnetic charge and the angular momentum respectively,
and $u=\sfrac{\tilde{P}\tilde{Q}}{G_4J}$.
\correctedA{
The dilaton and the gauge field are regular.
For the concrete expression of them,}
see \cite{Astefanesei:2006dd}.
Let us define $\phi' = \sqrt{u^2-1}\phi$.
Then, by taking a zero entropy limit $u\to 1$
with $J$ and $\phi'$ fixed to order one, the geometry turns out to be
\begin{align}
ds^2 = 2G_4 J \sin\theta
\left(
-r^2dt^2+\frac{dr^2}{r^2}+(d\phi'-rdt)^2\right)+
2G_4 J \sin\theta d\theta^2.
\end{align}
This form is exactly the one we found in the previous section and we can
see the AdS$_3$ structure in the first term.
Notice that, by U-duality, this black hole is related to a broad class
of extremal black holes appearing in string theory. Therefore, we can say
AdS$_3$ structure
emerges in the zero entropy limit of them, too.

Extremally rotating NS5-brane is an another example and
we can also see the emergence of AdS$_3$ structure in the
zero entropy limit\cite{Nakayama:2008kg}.

\subsection{Vacuum 5D black holes}
\label{Vacuum_5D_Black_Holes}

For the purpose above,
we consider 5D pure Einstein gravity 
with zero cosmological constant and then analyze
vacuum 5D extremal black holes with two $\mathrm{U}(1)$ symmetries. 
After Kaluza-Klein reduction along a $\mathrm{S}^1$ fiber
\correctedA{and switching to the Einstein frame},
these black holes reduce to 
the extremal black holes in 4D Einstein-Maxwell-dilaton gravity discussed above.
Here we concentrate on the cases with zero cosmological constant
for simplicity, but generalization to the case with cosmological constant
is straightforward.

When $\mathrm{SL}(2,\R)\times \mathrm{U}(1)^2$ symmetry is assumed,
the explicit form of the near horizon geometry of these black holes is
classified by \cite{Kunduri:2008rs}.
As for slow rotating
black holes whose angular momenta are bounded from above and the topology 
of horizon is $\mathrm{S}^3$, it is written as  
\begin{align}
ds^2 = \frac{\Gamma(\sigma)}{c_0^2}
\left(
-r^2dt^2+\frac{dr^2}{r^2}\right)+\frac{\Gamma(\sigma)}{Q(\sigma)}d\sigma^2
+\gamma_{ij}\left(dx^i-\frac{\bar{k}_{x^i}}{c_0^2}r dt\right)
\left(dx^j-\frac{\bar{k}_{x^j}}{c_0^2}rdt\right),
\end{align}
where 
\begin{align}
Q(\sigma) &= -c_0^2\sigma^2+c_1\sigma +c_2, \\
\gamma_{ij}dx^i dx^j &= 
\frac{P(\sigma)}{\Gamma(\sigma)}\left(
dx^{1}+\frac{\sqrt{-c_1c_2}}{c_0 P(\sigma)}dx^2
\right) ^2 +\frac{Q(\sigma)}{P(\sigma)}(dx^2)^2, \\
P(\sigma) &= c_0^2\sigma^2-c_2, \\
\Gamma &= \sigma,
\end{align}
and $\bar{k}_{x^1} =1$, $\bar{k}_{x^2}=0$.  
Parameters satisfy $c_1 >0$, $c_2<0$ and $c_0>0$ and
$\sigma$ takes $\sigma_1 <\sigma <\sigma_2$, 
where $\sigma_1$ and $\sigma_2$ are roots of $Q(\sigma)=0$. 
Explicit forms of these quantities are  written as 
\begin{align}
\sigma_{1} = \frac{1}{2}
\frac{c_1}{c_0^2}\left(1-\sqrt{1+\frac{4c_2c_0^2}{c_1^2}}
\right), \qquad
\sigma_{2} =
\frac{1}{2}
\frac{c_1}{c_0^2}\left(1+\sqrt{1+\frac{4c_2c_0^2}{c_1^2}}
\right).
\end{align}
For example,
\correctedA{Myers-Perry black hole and the slow rotating Kaluza-Klein black hole}
have this near horizon geometry 
in the extreme as we will explain. As for the detailed relation 
between parameters, see Appendix \ref{Some_Relations}.  

In order to make the regularity of the geometry manifest, 
we apply coordinate transformation so that new coordinates $\phi_1,\phi_2$ 
have periodicity $2\pi$: 
\begin{align}
x^1 = \frac{2\sqrt{-c_2}}{c_0^3(\sigma_2-\sigma_1)}(\phi_1-\phi_2), \qquad
x^2 = -\frac{2\sqrt{c_1}}{c_0^2(\sigma_2-\sigma_1)}(\sigma_1\phi_1-\sigma_2\phi_2).
\end{align}   
Then the corresponding parameter $\bar{k}_{\phi_1},\bar{k}_{\phi_2}$ are determined by a relation 
$\bar{k}=\bar{k}_{x^1}\partial_{x^1}+\bar{k}_{x^2}\partial_{x^2}
=\bar{k}_{\phi_1}\partial_{\phi_1}+\bar{k}_{\phi_2}\partial_{\phi_2}$,  
and the explicit forms are 
\begin{align}
\bar{k}_{\phi_1} = \frac{c_0^3}{2\sqrt{-c_2}}\sigma_2, \qquad
\bar{k}_{\phi_2} = \frac{c_0^3}{2\sqrt{-c_2}}\sigma_1.
\end{align}
By using the new coordinate, $\gamma_{ij}$ is written as 
\begin{align}
\gamma_{ij}dx^idx^j
 = f(\sigma)(d\phi_1)^2+2g(\sigma)d\phi_1 d\phi_2 + h(\sigma)(d\phi_2)^2,
\end{align}
where 
\begin{align}
f(\sigma) &= 
\frac{-4c_2}{c_0^6(\sigma_2-\sigma_1)^2}\frac{P}{\sigma}
\left(1-\frac{c_1\sigma_1}{P}\right)^2
+\frac{Q}{P}\frac{4c_1}{c_0^4(\sigma_2-\sigma_1)^2}\sigma_1^2, \\
g(\sigma) &= \frac{4c_2}{c_0^6(\sigma_2-\sigma_1)^2}\frac{P}{\sigma}
\left(1-\frac{c_1\sigma_1}{P}\right)
\left(1-\frac{c_1\sigma_2}{P}\right)-\frac{Q}{P}
\frac{4c_1}{c_0^4(\sigma_2-\sigma_1)^2}\sigma_1\sigma_2,
\\
h(\sigma) &= \frac{-4c_2}{c_0^6(\sigma_2-\sigma_1)^2}\frac{P}{\sigma}
\left(1-\frac{c_1\sigma_2}{P}\right)^2 +\frac{Q}{P}
\frac{4c_1}{c_0^4(\sigma_2-\sigma_1)^2}\sigma_2^2. 
\end{align}
Then the total metric is  
\begin{align}
ds^2 &= \frac{\sigma}{c_0^2}
\Bigg[
-r^2dt^2+\frac{dr^2}{r^2} +\frac{c_0^2}{Q}d\sigma^2 
+\frac{c_0^2}{\sigma}\frac{fh-g^2}{h}\left(
d\phi_1-k_{\phi_1}rdt
\right)^2 \nonumber \\ 
& \qquad\qquad\qquad\qquad
+\frac{c_0^2h}{\sigma}\left(
d\phi_2+\frac{g}{h}d\phi_1 -\left(k_{\phi_2}+\frac{g}{h}k_{\phi_1}\right)rdt
\right)^2 
 \Bigg], 
 \label{standard}
\end{align}
where
\begin{align}
k_{\phi_1} =\frac{\bar{k}_{\phi_1}}{c_0^2} = \frac{c_0}{2\sqrt{-c_2}}\sigma_2, \qquad
k_{\phi_2} =\frac{\bar{k}_{\phi_2}}{c_0^2} = \frac{c_0}{2\sqrt{-c_2}}\sigma_1.
\end{align}
This is the ``standard form'' introduced in the previous section under the identification 
\begin{align}
&\qquad A(\theta)^2 = \frac{\sigma}{c_0^2}, \qquad B(\theta)^2 = \frac{c_0^2}{\sigma}\frac{fh-g^2}{h},
\label{5Dvac_AB}
\\
&C(\theta)^2 = \frac{c_0^2h}{\sigma},\qquad D(\theta)^2 = \frac{g}{h}, 
\qquad F(\theta)^2d\theta^2 = \frac{d\sigma^2}{Q(\sigma)}.
\label{5Dvac_CDF}
\end{align} 

As we explained the near horizon geometry, we consider the 
zero entropy limit of it. For this purpose, let us write down the expression 
of the entropy and Frolov-Thorne temperatures corresponding to $\phi_1$-cycle and 
$\phi_{2}$-cycle: 
\begin{align}
&S_{\mathit{BH}} = \frac{4\pi^2}{4G_5}\int^{\sigma_2}_{\sigma_1} d\sigma \sqrt{\frac{\sigma(hf-g^2)}{Q}}, \\
&T_{\phi_1} = \frac{1}{2\pi k_{\phi_1}},\qquad\qquad T_{\phi_2} =\frac{1}{2\pi k_{\phi_2}}.
\end{align}

According to the discussion of previous section, zero entropy limit 
corresponds to $hf-g^2 = 0$.  
As a nontrivial example, we consider $c_2\to 0$. 
This corresponds to the zero entropy limit of \correctedA{the} extremal Myers-Perry black hole 
as can be seen by using the list of identification in Appendix \ref{Some_Relations}. 
By expanding with respect to $c_2$, we have
\begin{align}
\sigma_{1}=- \frac{c_2}{c_1}+\mathcal{O}(c_2^2), \qquad \sigma_2 = \frac{c_1}{c_0^2}
+\frac{c_2}{c_1}+\mathcal{O}(c_2^2), 
\label{5Dvac_MPtype_sigma12}
\end{align}
and then 
\begin{align}
k_{\phi_1} = \frac{c_1}{2c_0\sqrt{-c_2}}-\frac{c\sqrt{-c_2}}{2c_1}+\mathcal{O}\left((-c_2)^{3/2}\right), 
\qquad k_{\phi_2} =\frac{c_0\sqrt{-c_2}}{2c_1}+\mathcal{O}\left((-c_2)^{3/2}\right).
\label{5Dvac_MPtype_k12}
\end{align}
Moreover, since 
\begin{align}
f(\sigma) &= -\frac{4\sigma}{c_1^2}c_2 +\frac{4(c^2\sigma-c_1)}{c_1^4}c_2^2+\mathcal{O}(c_2^3),
\label{5Dvac_MPtype_f}
\\ 
g(\sigma) &= \frac{4(c_0^2\sigma-c_1)}{c_0^2c_1^2}c_2+\mathcal{O}(c_2^2),
\label{5Dvac_MPtype_g}
\\
h(\sigma) &= -\frac{4c_1(c_0^2\sigma-c_1)}{c_0^6\sigma}+\mathcal{O}(c_2),  
\label{5Dvac_MPtype_h}
\end{align}
we have 
\begin{align}
\frac{fh-g^2}{h} &= -\frac{4\sigma}{c_1^2}c_2 + \frac{4(c_0^4\sigma^2+3c_0^2c_1\sigma-c_1^2)}{c_1^5}c_2^2
+\mathcal{O}(c_2^3),
\label{5Dvac_MPtype_B}
\\
\frac{g}{h} &= -\frac{C^4\sigma}{c_1^3}c_2+\mathcal{O}(c_2^2). 
\label{5Dvac_MPtype_D}
\end{align}

\correctedA{
Now from \eqref{5Dvac_AB}\eqref{5Dvac_CDF}\eqref{5Dvac_MPtype_sigma12}%
\eqref{5Dvac_MPtype_k12}\eqref{5Dvac_MPtype_h}%
\eqref{5Dvac_MPtype_B}\eqref{5Dvac_MPtype_D},
it is easily shown that all of the regularity conditions
\eqref{5d_BC_limit}\eqref{5d_Bk1_limit}\eqref{5d_Dk2_limit}
are indeed satisfied in the current limit,
with $\epsilon\sim\sqrt{-c_2}$.
}
In order to
\correctedA{see the regularity manifestly}, 
we rescale the radial coordinate $r=k_{\phi_1}r'$, rewrite the 
metric by using $r'$ and then take $c_2\to 0$ limit. 
Explicitly, the metric turns out to be a regular form
\begin{align}
ds^2 &= \frac{c_0^4\sigma^2+c_1c_0^2\sigma-c_1^2}{4c_0^4c_1}r'^2dt^2+\frac{\sigma}{c_0^2}\left(
-2r'dtd\phi_1+\frac{dr'^2}{r'^2}
\right) \nonumber \\
& \qquad\qquad\qquad
+ \frac{1}{(c_1-c_0^2\sigma)}d\sigma^2 
+\frac{4c_1(c_1-c_0^2\sigma)}{c_0^6\sigma}
\left(d\phi_2-\frac{c^2}{4c_1}\sigma r' dt\right)^2.
\end{align}
Here, due to some annoying terms, AdS$_3$ factor does not appear but 
the regularity is manifest. 
The similar situation occurs in the setup of \cite{Guica:2010ej} when the 
radial coordinate is rescaled as above.

Let us next return to the coordinate \eqref{standard} and     
introduce $\psi=\phi_1, \phi= \phi_2$ corresponding to 
the angular variables of Myers-Perry black hole (see Appendix \ref{Some_Relations}). 
By regarding $\correctedA{d}\psi'=\correctedA{d}\psi/k_{\phi_1}$ and $\correctedA{d}\phi$ as order one 
quantities and taking $c_2\to 0$, we obtain the metric in the zero entropy limit as 
\begin{align}
ds^2 = 
 \frac{\sigma}{c_0^2}\left(
-r^2dt^2+\frac{dr^2}{r^2}+(d\psi'-rdt)^2\right) +\frac{1}{(c_1-c_0^2\sigma)}d\sigma^2
+\frac{4c_1(c_1-c_0^2\sigma)}{c_0^6\sigma}d\phi^2.
\label{mp_zero}
\end{align}
Here $0\leq \sigma \leq c_1/c_0^2$, $\phi\sim\phi+2\pi$ and $\psi'$-cycle shrink 
to zero, as explained in the previous section.  
Therefore, up to the $\sigma$-dependent overall factor, the first term is 
AdS$_3$ with vanishing periodicity in $\psi'$ direction. 

As for the zero entropy limit, there is an another possibility 
corresponding to \correctedA{the} extremal slow rotating Kaluza-Klein black hole. 
We first introduce $L$ as a sufficiently large quantity and consider 
a limit 
\begin{align}
c_0^2 = c_0'^2L,\qquad c_1 = c_1'L^2, \qquad c_2 = c_2' L \qquad (L\to \infty),   
\end{align}
where quantities with prime are order one. 
In this case, up to leading order in $1/L$ expansion,
\begin{align}
\sigma_1 = -\frac{c'_2}{c'_1}L^{-1}, \qquad \sigma_2 = \frac{c'_1}{2c_0'^2}L.
\end{align}
and 
\begin{align}
k_{\phi_1} =\frac{c'_1}{4c_0'\sqrt{-c'_2}}L,\qquad k_{\phi_2} = \frac{\sqrt{-c'_2}c_0'}{2c'_1}L^{-1}.
\label{5Dvac_KKtype_k12}
\end{align}
By introducing $\sigma' = \sigma /L$, we also have    
\begin{align}
f = -\frac{16c'_2}{c'_1{}^2}\sigma L^{-2}, \quad
g = \frac{8c'_2(2c'_0{}^2\sigma' -c'_1)}{c'_1{}^2 c_0'{^2}}L^{-2}, \quad
h = -\frac{4c'_1(c_0'{}^2\sigma'-c'_1)}{c_0'{}^6\sigma'},
\label{5Dvac_KKtype_fgh}
\end{align}
in the leading order.
\correctedA{
Therefore, also in this limit, we can make sure that
\eqref{5Dvac_AB}\eqref{5Dvac_CDF}\eqref{5Dvac_KKtype_k12}\eqref{5Dvac_KKtype_fgh}
satisfy
\eqref{5d_BC_limit}\eqref{5d_Bk1_limit}\eqref{5d_Dk2_limit},
with $\epsilon\sim L^{-1}$.
}

Let us next introduce a new coordinate  
\begin{align}
\phi = \phi_1 - \phi_2, 
\qquad y = 2\tilde{P}(\phi_2+\phi_1),  
\end{align}
corresponding to the angular variables of \correctedA{the} extremal 
slow rotating  Kaluza-Klein black hole,
and set $\phi\sim\phi+2\pi$ and $y\sim y+8\pi\tilde{P}$,
before taking the zero entropy limit. 
Here $\tilde{P}$ is the magnetic charge of the Kaluza-Klein 
black hole.
Detailed relations to the extremal slow rotating Kaluza-Klein 
black hole are summarized in Appendix \ref{Some_Relations}. 

Then by regarding $\correctedA{d}\phi'=\correctedA{d}\phi/k_{\phi}=\correctedA{d}\phi/(k_{\phi_1}-k_{\phi_2})$ and $\correctedA{d}y-2\tilde{P}\correctedA{d}\phi\,(=4\tilde{P}\correctedA{d}\phi_2)$ as order 
one quantities and taking the zero entropy limit as above, we have   
\begin{align}
ds^2 &= \frac{\sigma'}{c_0'^2}
\left(
-r^2dt^2 +\frac{dr^2}{r^2}
+\left(d\phi'-rdt 
\right)^2
\right) \nonumber \\
& \qquad\qquad\quad
+\frac{1}{c'_1-c_0^2\sigma'}d\sigma'{}^2
+\frac{1}{16\tilde{P}^2}\frac{4c'_1(c'_1-c_0'{}^2\sigma')}{c_0'{}^6\sigma'}
\left(dy-2\tilde{P}d\phi \right)^2.  
\label{kk_zero}
\end{align} 
Again, the period of $\phi'$ shrinks to zero in this limit.%
\footnote{
In this case
in the zero entropy limit, actually we encounter
another singularity than the one due to the shrink of $\mathrm{S}^1$-cycle
in the AdS$_3$ factor. It appears at $\sigma=0$ for \eqref{mp_zero} and 
at $\sigma'=0$ for \eqref{kk_zero}.
}

\subsection{Black holes in 5D supergravity}
Next we consider with the black holes in 5D supergravity,
obtained from dimensional reduction of the rotating
D1-D5-P black holes in the compactified IIB supergravity.
They include the ones dealt in \cite{Guica:2010ej},
where they took the zero entropy limit from the fast rotating range.
Unlike \cite{Guica:2010ej},
we do not require $J_R=0$ and work on the purely 5D reduced theory
to clarify the emergence of the AdS$_3$,
because this system always has an AdS$_3$ structure
along the uplifted Kaluza-Klein direction.

We will take the zero entropy limit from the slow rotating range.
The 6D form of the near horizon metric of this system
was given in \cite{Azeyanagi:2008dk}%
\footnote{
The current authors called it ``very near horizon'' geometry there,
to distinguish it from the AdS$_3$ decoupling limit.
We redefined some signs from there to agree with our convention here.
}
as
\begin{align}
  ds_{(6)}^2
  =
  \frac{\lambda^2}{4}\bigg[
   &- r^2dt^2 +\frac{dr^2}{r^2}
   + 4r_+^2\big(dy-\frac{r}{2r_+}dt\big)^2
  \nonumber\\
  &
   + 4\bigg(
    d\theta^2
    + \sin^2\theta \Bigl(d\phi - \frac{2G_6}{\pi^2\lambda^4}J_{\psi}\,dy\Bigr)^2
    + \cos^2\theta \Bigl(d\psi - \frac{2G_6}{\pi^2\lambda^4}J_{\phi}\,dy\Bigr)^2
   \bigg)
  \bigg],
  \label{D1D5P_6D_metric}
\end{align}
where
\begin{align}
  0\le\theta&\le\frac{\pi}{2},\quad \phi \sim \phi+2\pi,\quad \psi \sim \psi+2\pi,\\
  \lambda^4&=Q_1Q_5,\\
  r_+ &= \frac{G_6}{\pi^3\lambda^4}S_{\mathit{BH}},
  \label{D1D5P_rplus}\\
  S_{\mathit{BH}} &= 2\pi\sqrt{\Big(\frac{\pi^2R}{2G_6}\Big)^2Q_1Q_5Q_p-J_{\phi}J_{\psi}},
  \label{D1D5P_entropy}\\
  Q_1,\;&\;Q_5,\;Q_p>0,\quad J_{\phi},\;J_{\psi}>0.
\end{align}
Here $G_6$ is 6D Newton constant, $R$ is the Kaluza-Klein radius,
$J_{\phi}$ and $J_{\psi}$ are angular momenta,
$Q_1$, $Q_5$, $Q_p$ are D1, D5 and Kaluza-Klein momentum charges respectively,
and $S_{\mathit{BH}}$ is the Bekenstein-Hawking entropy.
The zero entropy limit we want to take here is characterized
by cancellation of the two terms in the root in \eqref{D1D5P_entropy},
while keeping all the charges and angular momenta to be nonzero finite
together with $G_6$ and $R$.
{}From \eqref{D1D5P_rplus}, it immediately means the limit of
\begin{align}
  r_+ \to 0.
  \label{D1D5P_zeroentropy_limit}
\end{align}

After some short algebra, the metric \eqref{D1D5P_6D_metric} can be rewritten as
\begin{align}
  ds_{(6)}^2
  = \frac{\lambda^2}{4}
   \bigg[
    - r^2dt^2 +\frac{dr^2}{r^2}
    + B(\theta)^2e_{\phi}^2
    + C(\theta)^2(e_{\psi} + D(\theta)e_{\phi})^2
   \bigg]
   + \lambda^2d\theta^2
   + \Phi(\theta)^2\Big(dy - \mathcal{A}\Big)^2,
\end{align}
where
\begin{align}
  \Phi(\theta)^2 &= \frac{4G_6^2(J_{\phi}^2\cos^2\theta + J_{\psi}^2\sin^2\theta)}{\pi^4\lambda^6}+\lambda^2r_+^2,
  \label{D1D5P_dilaton}\\
  \mathcal{A} &= k_yrdt \correctedA{+} \frac{2\pi^2G_6\lambda^4}{4G_6^2(J_{\phi}^2\cos^2\theta+J_{\psi}^2\sin^2\theta)+\pi^4\lambda^8r_+^2}
   \big(J_{\psi}\sin^2\theta\,e_{\phi} + J_{\phi}\cos^2\theta\,e_{\psi}\big),
   \label{D1D5P_5D_gaugefield}\\
   B(\theta)^2 &= \frac{4\pi^4\lambda^8r_+^2\sin^2\theta}{4G_6^2J_{\psi}^2\sin^2\theta + \pi^4\lambda^8r_+^2},
   \label{D1D5P_B}\\
   C(\theta)^2 &= \frac{4(4G_6^2J_{\psi}^2\sin^2\theta+\pi^4\lambda^8r_+^2)\cos^2\theta}{4G_6^2(J_{\phi}^2\cos^2\theta+J_{\psi}^2\sin^2\theta)+\pi^4\lambda^8r_+^2},
   \label{D1D5P_C}\\
   D(\theta) &= \frac{-4G_6^2J_{\phi}J_{\psi}\sin^2\theta}{4G_6^2J_{\psi}^2\sin^2\theta + \pi^4\lambda^8r_+^2},
   \label{D1D5P_D}\\
  e_{\phi} &= d\phi - k_{\phi}rdt,\quad e_{\psi} = d\psi - k_{\psi}rdt,\\
   k_{\phi}&=\frac{G_6J_{\psi}}{\pi^2\lambda^4r_+},
   \quad k_{\psi}=\frac{G_6J_{\phi}}{\pi^2\lambda^4r_+},
   \quad k_y=\frac{1}{2r_+}.
   \label{D1D5P_k}
\end{align}

Because the $y$-cycle never shrinks or blows up from \eqref{D1D5P_dilaton},
we can safely reduce the system into 5D theory,
with the dilatonic field $\Phi(\theta)$,
Kaluza-Klein gauge field $\mathcal{A}$, 5D metric
\begin{align}
  ds_{(5)}^2
  = \frac{\lambda^2}{4}
   \bigg[
    - r^2dt^2 +\frac{dr^2}{r^2}
    + B(\theta)^2e_{\phi}^2
    + C(\theta)^2\big(e_{\psi} + D(\theta)e_{\phi}\big)^2
   \bigg]
   + \lambda^2d\theta^2,
   \label{D1D5P_5D_metric}
\end{align}
and the Newton constant being $G_5=\sfrac{G_6}{2\pi R}$.
\correctedA{
Note that the gauge field $\mathcal{A}$ \eqref{D1D5P_5D_gaugefield} is finite in this limit,
although it is obvious from the finiteness of \eqref{D1D5P_6D_metric}.
}
Manifestly \eqref{D1D5P_5D_metric} has the very form of
\eqref{5d_ansatz}, and in the limit \eqref{D1D5P_zeroentropy_limit},
the behaviors of 
\correctedA{
$B(\theta)$, $C(\theta)$, $D(\theta)$ and $k_1$, $k_2$
in \eqref{D1D5P_B}\eqref{D1D5P_C}\eqref{D1D5P_D}\eqref{D1D5P_k}
do satisfy the conditions
\eqref{5d_BC_limit}\eqref{5d_Bk1_limit}\eqref{5d_Dk2_limit},
}
under an identification $\epsilon=r_+$.
In fact, in the limit the 5D metric goes to just the form of
\eqref{5d_AdS3_manifest},
with
\begin{align}
  A(\theta) &= \frac{\lambda}{2},
  \quad
  F(\theta) = \lambda,\\
  C(\theta)^2 &= \frac{4J_{\psi}^2\sin^2\theta\cos^2\theta}{J_{\phi}^2\cos^2\theta+J_{\psi}^2\sin^2\theta},\\
  \phi'_1 &= \frac{\phi}{k_{\phi}},
  \quad \phi'_2 = \psi + D\phi,
  \quad k_{\phi}=\frac{-2RG_5J_{\psi}}{\pi\lambda^4\epsilon},
  \quad D = -\frac{J_{\phi}}{J_{\psi}}
  \label{D1D5P_AdS3_angular_coordinates}.
\end{align}
This is a very special case of emergent AdS$_3$,
in that $A(\theta)$ is constant and so the AdS$_3$ is not fibered.%
\footnote{\correctedA{
The metric is written in the string frame here.
In the Einstein frame, $A(\theta)$ has a $\theta$-dependence
unless $J_{\phi}=J_{\psi}$.
}}
Furthermore, we see from \eqref{D1D5P_AdS3_angular_coordinates} that,
if $J_{\phi}$ is $J_{\psi}$ times an integer,
$\phi'_2$ has a proper periodicity as we discussed in
\S\ref{5d_general_AdS3_emergence}.
In that case the spacetime is a direct product of an
(infinitesimally orbifolded) AdS$_3$ and a squashed $\mathrm{S}^2$.
Finally, the most special case is $J_{\phi}=J_{\psi}$,
that is, zero-entropy BMPV.
In this case $C(\theta)=\sin 2\theta$
and so we obtain a non-squashed
$\mathrm{AdS}_3\times \mathrm{S}^2$,
as was seen in \cite{Guica:2007gm,Guica:2010ej}.

\section{Implications to the Kerr/CFT}
\label{Implications_to_KerrCFT}

In this section we shortly consider the correspondence between the
central charges of the Kerr/CFT and AdS$_3$/CFT$_2$
in the zero entropy limit.
For simplicity we work on the 4D case here,
but 5D case is almost the same.

\subsection{Kerr/CFT in the zero entropy limit}

Let us consider again the 4D metric \eqref{4d_ansatz},
\begin{align}
  ds^2 = A(\theta)^2
\correctedA{
  \left[
    -r^2dt^2+\frac{dr^2}{r^2}+B(\theta)^2\big(d\phi-krdt\big)^2
  \right]
}
  +F(\theta)^2d\theta^2.
\end{align}
First of all, for clarification of the discussions below,
we transform this metric from Poincar\'e
form to the global form,
\begin{align}
  ds^2 = A(\theta)^2 \left[-(1+\tilde{r}^2)d\tilde{t}^2
    +\frac{d\tilde{r}^2}{1+\tilde{r}^2}
    +B(\theta)^2\big(d\tilde{\phi}-k\tilde{r}d\tilde{t}\big)^2\right]
   +F(\theta)^2d\theta^2.
   \label{4d_global_metric}
\end{align}
This coordinates transformation $(t,r,\phi)\to (\tilde{t},\tilde{r},\tilde{\phi})$
can be carried out without any change on the forms and values of
$A(\theta)$, $B(\theta)$, $F(\theta)$ and $k$.
For this geometry, the usual Kerr/CFT procedure with the chiral Virasoro generators
\begin{align}
  \xi_n = -inre^{-in\tilde{\phi}}\partial_{\tilde{r}} - e^{-in\tilde{\phi}}\partial_{\tilde{\phi}},
\end{align}
gives the central charge
\begin{align}
  c_{\mathrm{Kerr/CFT}} = \frac{3k}{G_4} \int\!\!d\theta\, A(\theta)B(\theta)F(\theta).
  \label{4d_kerrcft_center}
\end{align}
\correctedB{
The corresponding Frolov-Thorne temperature is
\begin{align}
  T_{FT} = \frac{1}{2\pi k},
\end{align}
as usual.
Therefore the entropy is
\begin{align}
  S_{\mathrm{Kerr/CFT}}
  = \frac{\pi^2}{3}c_{\mathrm{Kerr/CFT}}T_{FT}
  = \frac{\pi}{2G_4}\int\!\!d\theta\, A(\theta)B(\theta)F(\theta),
  \label{4d_kerrcft_entropy}
\end{align}
which of course agrees with $S_{BH}$ \eqref{4d_entropy}.
In particular, 
near the zero entropy limit \eqref{4d_zeroentropy_limit},
the central charge  \eqref{4d_kerrcft_center} is expanded as
}
\begin{align}
  c_{\mathrm{Kerr/CFT}} = \frac{3}{G_4}\int\!\!d\theta\, A(\theta)F(\theta)
  \correctedB{+ \mathcal{O}\Big(\frac{1}{k^2}\Big)},
  \label{4d_zeroentropy_kerrcft_center}
\end{align}
\correctedB{
and the entropy \eqref{4d_kerrcft_entropy} becomes
\begin{align}
  S_{\mathrm{Kerr/CFT}}
  =\frac{\pi}{2kG_4}\int\!\!d\theta\, A(\theta)F(\theta)
  + \mathcal{O}\Big(\frac{1}{k^3}\Big).
  \label{4d_zeroentropy_kerrcft_entropy}
\end{align}}

\subsection[AdS$_3$/CFT$_2$ in the emergent AdS$_3$]%
{AdS$_{\bm{3}}$/CFT$_{\bm{2}}$ in the emergent AdS$_3$}

\correctedB{
In the last subsection, we examined 
the Kerr/CFT
near the zero entropy limit.
In this limit,}
the metric \eqref{4d_global_metric} itself becomes
\begin{align}
  ds^2 &=
  A(\theta)^2 
\correctedA{
  \left[
    -(1+\tilde{r}^2)d\tilde{t}^2+
    \frac{d\tilde{r}^2}{1+\tilde{r}^2}
    +(d\tilde{\phi}'-\tilde{r}d\tilde{t})^2
  \right]
}
  + F(\theta)^2d\theta^2,
  \label{4d_global_AdS3_A}\\
  \tilde{\phi}'&=\frac{\tilde{\phi}}{k},
  \quad \tilde{\phi}'\sim\tilde{\phi}'+\frac{2\pi}{k},
  \quad \correctedB{(k\to\infty)}
  \label{4d_global_phi_periodicity}
\end{align}
in exactly the same way as \eqref{4d_AdS3}.
\correctedB{
Since an AdS$_3$ structure is included,
it is expected that this geometry has a non-chiral dual theory
and the Kerr/CFT above is a chiral part of it.%
\footnote{\correctedB{
Chiral CFT$_2$ and an AdS$_3$ structure were discussed in \cite{Balasubramanian:2009bg}.
}}
We will partly demonstrate it below.}

\correctedB{
First we stress again that the AdS$_3$ is orbifolded by an infinitesimally
narrow period \eqref{4d_global_phi_periodicity}.}
In this form of the metric, if we \correctedB{change}
the coordinates range as
\begin{align}
  \tilde{t}&\sim\tilde{t}+4\pi,
  \quad \correctedB{-\infty<\tilde{r}<\infty},
  \quad -\infty<\tilde{\phi}<\infty,
  \label{4d_global_AdS3_coordinatesA_range}
\end{align}
they cover the whole AdS$_3$ as a hyperbolic hypersurface in $\R^{2,2}$.%
\footnote{
\correctedB{
The boundary of $\mathrm{AdS}_3$
corresponds to $\tilde{r}=\pm\infty$ or $\tilde{\phi}=\pm\infty$.
It is a special property of the unorbifolded case.
When $\tilde{\phi}$ has a period,
the $\tilde{r}=-\infty$ region is identified with the $\tilde{r}=\infty$ region,
and so the boundary is described simply by $\tilde{r}=\infty$.
}}
Even under the replaced periodicity of $\tilde{\phi}$ in
\eqref{4d_global_AdS3_coordinatesA_range},
the results for the central charge of the Kerr/CFT
\eqref{4d_kerrcft_center} 
\correctedB{\eqref{4d_zeroentropy_kerrcft_center}}
remain true.
Generally speaking,
multiplying the period of the S$^1$ coordinate
by $n$ alters the corresponding
Frolov-Thorne temperature $T_{\mathit{FT}}=\sfrac{1}{2\pi k}$
to $T_{\mathit{FT}}'=\sfrac{T_{\mathit{FT}}}{n}$,
but leaves the central charge unchanged.
\correctedB{%
It is consistent with holography, because the central charges
are local quantities in the dual CFT and so should not depend on
the periodicity.}

\correctedB{From this observation, 
we carry out a coordinates transformation and map}
the metric \eqref{4d_global_AdS3_A}
from the form of an S$^1$ fibered AdS$_2$
to a more conventional AdS$_3$ form,
\begin{align}
  ds^2 = 4A(\theta)^2\left[-(1+\rho^2)d\tau^2 + \frac{d\rho^2}{1+\rho^2} + \rho^2d\psi^2\right] + F(\theta)^2d\theta^2.
  \label{4d_global_AdS3_B}
\end{align}
Corresponding to the coordinates range \eqref{4d_global_AdS3_coordinatesA_range} for \eqref{4d_global_AdS3_A},
this coordinate system \eqref{4d_global_AdS3_B}
again covers the whole AdS$_3$
as a hyperbolic hypersurface, when we take 
\begin{align}
  \tau&\sim\tau+2\pi,\quad \rho\ge 0,\quad \psi\sim\psi+2\pi.
  \label{4d_global_AdS3_coordinatesB_range}
\end{align}
Now in the coordinate system $(\tau,\rho,\psi,\theta)$,
we can adopt, as the asymptotic symmetry generators for \eqref{4d_global_AdS3_B},
the Virasoro generators obtained in \cite{Brown:1986nw},
or those recently proposed in \cite{Porfyriadis:2010vg},
\begin{align}
  \xi^{(R)}_n&=\frac{1}{2}
   \Big(e^{in(\correctedB{\tau}+\correctedA{\psi})}\partial_{\correctedB{\tau}}
   - inre^{in(\correctedB{\tau}+\correctedA{\psi})}\partial_{\correctedB{\rho}}
   + e^{in(\correctedB{\tau}+\correctedA{\psi})}\partial_{\correctedA{\psi}}\Big),\\
  \xi^{(L)}_n&=\frac{1}{2}
   \Big(e^{in(\correctedB{\tau}-\correctedA{\psi})}\partial_{\correctedB{\tau}}
   - inre^{in(\correctedB{\tau}-\correctedA{\psi})}\partial_{\correctedB{\rho}}
   - e^{in(\correctedB{\tau}-\correctedA{\psi})}\partial_{\correctedA{\psi}}\Big).
\end{align}
By explicit calculation, both choices lead to the same result
\begin{align}
  c^{(R)} = c^{(L)}
  = \frac{3}{G_4}\int\!\!d\theta\, A(\theta)F(\theta).
  \label{4d_AdS3_central_charges}
\end{align}
This value exactly agrees with that of \eqref{4d_zeroentropy_kerrcft_center}.
It suggests that there are indeed some relations between the Kerr/CFT
and the AdS$_3$/CFT$_2$ coming from our emergent AdS$_3$.
Although they have been calculated under the periodicity \eqref{4d_global_AdS3_coordinatesB_range}, 
they are expected to be independent of it
for the same reason as above ---
the central charges are local quantities in the dual CFT.

{}From the periodicity \eqref{4d_global_phi_periodicity},
the Frolov-Thorne \correctedB{temperatures} of the system could be identified as
$T_{\correctedB{L}}=\sfrac{1}{2\pi k}$ \correctedB{and $T_{R}=0$},
as in \cite{Guica:2010ej}.
Then the entropy computed from Cardy formula is
\correctedA{
\begin{align}
  S_{\mathrm{AdS}_3}
  =\frac{\pi^2}{3}c^{(L)}T_{\correctedB{L}}
   \correctedB{+ \frac{\pi^2}{3}c^{(R)}T_{R}}
  =\frac{\pi}{2kG_4}\int\!\!d\theta\, A(\theta)F(\theta).
  \label{4d_S_AdS3}
\end{align}
\correctedB{It}
agrees with $S_{\mathrm{Kerr/CFT}}$
\correctedB{\eqref{4d_zeroentropy_kerrcft_entropy}},
up to $\mathcal{O}(1/k^3)$ correction terms.
This result is reasonable, or better than expected.
For, from the first,
\eqref{4d_S_AdS3} is reliable only for leading order of $1/k$ expansion,
because the metric will be changed in higher order in $1/k$.
}

\section{Conclusions and Discussions}
\label{Conclusions_and_Discussions}

In this paper,
we studied the zero entropy limit for 
near horizon geometries of $D=4$ and $D=5$ general extremal black holes 
with $\mathrm{SL}(2,\R)\times \mathrm{U}(1)^{D-3}$ symmetry.
\correctedA{
We derived the conditions on the near horizon geometries of the
black holes in the zero entropy limit,
based on the expectation that they should remain regular.
Then we
}
found that they have AdS$_3$ structure in general,
although the periodicity shrinks to zero. 
\correctedA{We presented} some concrete examples, including extremal
5D Myers-Perry black hole,
5D Kaluza-Klein black hole and
black holes in 5D supergravity to see the emergence.
\correctedA{We also discussed} some relation between the chiral CFT$_2$ appearing in 
the Kerr/CFT and the non-chiral CFT$_2$ expected to be dual 
to AdS$_3$ emerging in the zero entropy limit.        

There are possible generalizations of our consideration to other setups.
For example,
generalization to higher dimensional cases would be valuable,
and finding more concrete examples would be also interesting.

Of course, there are many important points which should be addressed
in order to understand the Kerr/CFT correspondence
from the AdS$_3$ structures \correctedA{generally} investigated in this paper.
One of those which was not studied 
in this paper is what is an appropriate boundary condition.
In particular, it is totally \correctedA{unclear how}
such boundary condition will be changed under
the deformation to non-zero entropy black \correctedA{holes}.
We hope to return to these problems in near future.

\section*{Acknowledgements}
T.~A.~ and N.~O. are supported by the Japan Society for the Promotion of Science (JSPS).
S.~T.~ is partly supported by the Japan Ministry of Education, Culture, Sports, Science and Technology (MEXT).
This work is supported by the Grant-in-Aid for the Global COE program
``The Next Generation of Physics, Spun from Universality and Emergence''
from the MEXT. 
\vspace{10mm}

\appendix

\section{Derivation of the Regularity Conditions in 5D}
\label{5d_general_proof}

In this appendix, we give the detail of the derivation of
the regularity conditions explained in \S\ref{5d_general_zeroentropy_limit}.

Using $r'$ in \eqref{5d_zeroentropy_limit1},
the metric \eqref{5d_ansatz} is written as
\begin{align}
  ds^2 =
  A(\theta)^2
  \bigg[
   &\frac{B(\theta)^2k_1^2 + C(\theta)^2(k_2+D(\theta)k_1)^2 - 1}{\epsilon^2}r'^2dt^2
   + \frac{dr'^2}{r'^2}
 \nonumber\\
   &- \frac{B(\theta)^2k_1+C(\theta)^2D(\theta)(k_2+D(\theta)k_1)}{\epsilon}2r'dtd\phi_1
   - \frac{C(\theta)^2(k_2+D(\theta)k_1)}{\epsilon}2r'dtd\phi_2
 \nonumber\\
   &+ \big(B(\theta)^2+C(\theta)^2D(\theta)^2\big)d\phi_1^2
   + 2C(\theta)^2D(\theta)d\phi_1d\phi_2
   + C(\theta)^2d\phi_2^2
  \bigg]
  + F(\theta)^2d\theta^2
\end{align}
Each term here must not diverge, therefore,
from the finiteness of the
$dt^2$, $dtd\phi_1$, $dtd\phi_2$, $d\phi_1^2$, $d\phi_2^2$ terms respectively,
\begin{subequations}
\begin{align}
  B(\theta)^2k_1^2 + C(\theta)^2(k_2+D(\theta)k_1)^2 - 1 &= \mathcal{O}(\epsilon^2),
  \label{5d_dtdt_condition}\\
  B(\theta)^2k_1+C(\theta)^2D(\theta)(k_2+D(\theta)k_1) &= \mathcal{O}(\epsilon),
  \label{5d_dtdphi1_condition}\\
  C(\theta)^2(k_2+D(\theta)k_1) &= \mathcal{O}(\epsilon),
  \label{5d_dtdphi2_condition}\\
  B(\theta)^2+C(\theta)^2D(\theta)^2 &= \mathcal{O}(1),
  \label{5d_dphi1dphi1_condition}\\
  C(\theta)^2D(\theta) &= \mathcal{O}(1),
  \label{5d_dphi1dphi2_condition}\\
  C(\theta)^2 &= \mathcal{O}(1).
  \label{5d_dphi2dphi2_condition}
\end{align}
In particular, \eqref{5d_dphi1dphi1_condition} means
\begin{align}
  B(\theta)=\mathcal{O}(1),
  \quad C(\theta)D(\theta)=\mathcal{O}(1),
  \label{5d_dphi1dphi1_condition:2}
\end{align}
\label{5d_terms_conditions}
\end{subequations}
then \eqref{5d_dphi1dphi2_condition} is always satisfied
under \eqref{5d_dphi2dphi2_condition} and \eqref{5d_dphi1dphi1_condition:2}.

\subsection[Generality of ${B(\theta)\to 0,\; C(\theta)\sim 1}$]%
{Generality of $\bm{B(\theta)\to 0,\; C(\theta)\sim 1}$}

To achieve $B(\theta)C(\theta)\to 0$ while satisfying the above conditions
\eqref{5d_dphi2dphi2_condition} \eqref{5d_dphi1dphi1_condition},
there are three cases to be considered:
\begin{subequations}
\begin{align}
  B(\theta)&\to 0,\quad C(\theta)\sim 1.
  \label{5d_proof_BC_case_a}\\
  B(\theta)&\sim 1,\quad C(\theta)\to 0
  \label{5d_proof_BC_case_b}.\\
  B(\theta)&\to 0,\quad C(\theta)\to 0
  \label{5d_proof_BC_case_c}.
\end{align}
\end{subequations}
Here we will examine the cases of \eqref{5d_proof_BC_case_b} and \eqref{5d_proof_BC_case_c},
and show that they can be resulted in the case of \eqref{5d_proof_BC_case_a}.

\subsubsection[{$B(\theta)\sim 1,\; C(\theta)\to 0$}]%
{\bm{$B(\theta)\sim 1,\; C(\theta)\to 0$}}

This case is expressed with $\epsilon$ in \eqref{5d_zeroentropy_limit1} as
\begin{align}
  B(\theta)\sim 1,\quad C(\theta)\sim\epsilon.
\end{align}
Now \eqref{5d_dphi1dphi1_condition:2} leads to
\begin{align}
  B(\theta)^2 + C(\theta)^2D(\theta)^2 \sim 1,
  \quad C(\theta)^2D(\theta)=\mathcal{O}(\epsilon).
\end{align}
so after acting the swapping transformation $\mathcal{S}'$
\eqref{5d_Sp_transformation:2}, we obtain
\begin{align}
  \tilde{B}(\theta)\sim\epsilon,
  \quad \tilde{C}(\theta)\sim 1,
  \quad \tilde{D}(\theta)=\mathcal{O}(\epsilon). 
\end{align}
Therefore this is clearly the case of \eqref{5d_proof_BC_case_a},
especially with $D=0$.

\subsubsection[{$B(\theta)\to 0,\; C(\theta)\to 0$}]%
{\bm{$B(\theta)\to 0,\; C(\theta)\to 0$}}

In this case, we introduce new small parameters
$\epsilon_1$, $\epsilon_2$ and rewrite the condition as
\begin{align}
  B(\theta)\sim\epsilon_1,
  \quad C(\theta)\sim\epsilon_2,
  \quad \epsilon=\epsilon_1\epsilon_2,
  \quad \epsilon_1\to 0,
  \quad \epsilon_2\to 0.
  \label{5d_proof_BC_scaling_c}
\end{align}
With $\epsilon_1$ and $\epsilon_2$,
\eqref{5d_dtdphi1_condition},
\eqref{5d_dtdphi2_condition},\eqref{5d_dphi1dphi1_condition:2}
lead to
\begin{align}
  B(\theta)^2k_1 + C(\theta)^2D(\theta)(k_2+D(\theta)k_1) &= \mathcal{O}(\epsilon_1\epsilon_2),
  \label{5d_limit_condition_3.30}\\
  k_2 + D(\theta)k_1 &\sim \sfrac{\epsilon_1}{\epsilon_2},
  \label{5d_limit_condition_3.10}\\
  D(\theta) &= \mathcal{O}(\sfrac{1}{\epsilon_2}).
  \label{5d_limit_condition_3.20}
\end{align}
Notice that, since $d\phi_1d\phi_2$ and $d\phi_2^2$ terms go to zero in this case,
$dtd\phi_2$ term must not vanish for regularity and so
\eqref{5d_limit_condition_3.10} fixes the scaling order exactly,
rather than $\mathcal{O}(\sfrac{\epsilon_1}{\epsilon_2})$.
We see that,
from \eqref{5d_proof_BC_scaling_c} and \eqref{5d_limit_condition_3.10},
\begin{align}
  C(\theta)^2(k_2 + D(\theta)k_1)^2 \sim \epsilon_1^2,
\end{align}
therefore by \eqref{5d_dtdt_condition},
\begin{align}
  B(\theta)^2k_1^2 = 1 + \mathcal{O}(\epsilon_1^2),
\end{align}
which is equivalent to
\begin{align}
  B(\theta) = \frac{1}{k_1} + \mathcal{O}(\epsilon_1).
\end{align}
Thus $B(\theta)^2k_1\sim\epsilon_1\gg\epsilon_1\epsilon_2$,
so \eqref{5d_limit_condition_3.30} means
\begin{align}
   C(\theta)^2D(\theta)(k_2+D(\theta)k_1)
   = -B(\theta)^2k_1 + \mathcal{O}(\epsilon_1\epsilon_2)
   \sim\epsilon_1,
\end{align}
which leads to, by using \eqref{5d_limit_condition_3.10},
\begin{align}
  D(\theta)\sim \sfrac{1}{\epsilon_2}.
  \label{5d_proof_BC_scaling_c_orderof_D}
\end{align}

Now it is easy to see,
from \eqref{5d_proof_BC_scaling_c} and
\eqref{5d_proof_BC_scaling_c_orderof_D}, that
$\mathcal{S}'$ transforms $B(\theta)$, $C(\theta)$ and $D(\theta)$
into
\begin{align}
  \tilde{B}(\theta)\sim\epsilon_1\epsilon_2=\epsilon,
  \quad \tilde{C}(\theta)\sim 1,
  \quad \tilde{D}(\theta)\sim \epsilon_2.
\end{align}
Therefore the current case \eqref{5d_proof_BC_case_c}
is also transformed to the case of \eqref{5d_proof_BC_case_a},
again with $D=0$.

\subsection[Conditions under ${B(\theta)\to 0,\; C(\theta)\sim 1}$]%
{Conditions under $\bm{B(\theta)\to 0,\; C(\theta)\sim 1}$}
\label{5d_proof_B=0_case}

Under \eqref{5d_BC_limit},
\begin{align}
  B(\theta)\sim\epsilon,\quad C(\theta)\sim 1,
\end{align}
the finiteness conditions
\eqref{5d_dtdt_condition},\eqref{5d_dtdphi1_condition},\eqref{5d_dtdphi2_condition} and \eqref{5d_dphi1dphi1_condition:2} become, respectively,
\begin{align}
  B(\theta)^2k_1^2 + C(\theta)^2(k_2+D(\theta)k_1)^2 -1 &= \mathcal{O}(\epsilon^2),
  \label{5d_limit_condition_1.10}\\
  B(\theta)^2k_1 + C(\theta)^2D(\theta)(k_2+D(\theta)k_1) &= \mathcal{O}(\epsilon),
  \label{5d_limit_condition_1.20}\\
  k_2+D(\theta)k_1 &= \mathcal{O}(\epsilon),
  \label{5d_limit_condition_1.30}\\
  D(\theta) &= \mathcal{O}(1).
  \label{5d_limit_condition_1.35}
\end{align}
Then $C(\theta)^2(k_2+D(\theta)k_1)^2=\mathcal{O}(\epsilon^2)$
from \eqref{5d_limit_condition_1.30},
and so substituting it into \eqref{5d_limit_condition_1.10} yields
\begin{align}
  B(\theta)^2k_1^2 = 1 + \mathcal{O}(\epsilon^2),
\end{align}
or equivalently,
\begin{align}
  B(\theta)k_1 = 1 + \mathcal{O}(\epsilon^2),
  \label{5d_limit_condition_1.40}
\end{align}
where we can immediately see that
\begin{align}
  k_1\sim \frac{1}{\epsilon}.
  \label{5d_limit_condition_1.50}
\end{align}
Thus from \eqref{5d_limit_condition_1.30} and \eqref{5d_limit_condition_1.50},
\begin{align}
  D(\theta) = -\frac{k_2}{k_1} + \mathcal{O}(\epsilon^2),
\end{align}
which leads to, using
\eqref{5d_limit_condition_1.35} and \eqref{5d_limit_condition_1.50},
\begin{align}
  k_2 = \mathcal{O}(\epsilon^{-1}).
\end{align}
Therefore we have successfully shown
\eqref{5d_Bk1_limit} and \eqref{5d_Dk2_limit}
to be the regularity condition for the geometry
in the zero entropy limit.

\section{Some Relations}
\label{Some_Relations}

In this appendix we summarize the relations between the parameters 
$c_1$, $c_2$, $c_0^2$ appearing in \S\ref{Vacuum_5D_Black_Holes}
and physical quantities in the extremal Myers-Perry black hole
and \correctedA{the} extremal slow rotating Kaluza-Klein black hole\cite{Kunduri:2008rs}.

\subsection{Myers-Perry black hole}
The \correctedA{relation} to \correctedA{Myers-Perry black hole is} as follows. 
Let us first use a scaling symmetry 
\begin{align}
c_0^2\to \lambda c_0^2,\qquad c_1\to \lambda^2 c_1,\qquad c_2\to \lambda^3 c_2,\qquad
x_1 \to \lambda^{-1}x_1,
\end{align}
to set $c_0^2 =c_1$ and introduce  
\begin{align}
a = \frac{1}{\sqrt{c_1}}+\frac{\sqrt{c_1+4c_2}}{c_1}, 
\qquad b = \frac{1}{\sqrt{c_1}}-\frac{\sqrt{c_1+4c_2}}{c_1}. 
\end{align}
By changing coordinates  
\begin{align}
\cos^2\theta = \frac{\sigma-\sigma_1}{\sigma_2-\sigma_1},  \quad
x^1 = -\frac{\sqrt{ab}(a+b)^2}{2(a-b)}(-\psi+\phi),\quad 
x^2 = \frac{a+b}{a-b}(-b\psi+a\phi),
\end{align}
so that $0\leq\theta\leq \pi/2$, $\phi\sim \phi+2\pi$ and $\psi\sim \psi+2\pi$. 
The horizon data are summarized as follows:
\begin{align}
\gamma_{ij}dx^idx^j &= \frac{1}{\rho_+^2}
\Big[(r_+^2+a^2)^2\sin^2\theta d\phi^2 + (r_+^2+b^2)^2\cos^2\theta d\psi^2 
\Big] \nonumber \\\ &  \qquad\qquad 
+ \frac{1}{r_+^2\rho_+^2}\Big[
b(r_+^2+a^2)\sin^2\theta d\phi +a(r_+^2+b^2)\cos^2\theta d\psi
\Big]^2, \\ 
\frac{\sigma}{Q(\sigma)}d\sigma^2  &= \rho_+^2d\theta^2, \qquad
\Gamma = \frac{\rho_+^2 r_+^2}{(r_+^2+a^2)(r_+^2+b^2)}, \\ 
\bar{k}_{\phi} &= \frac{2a r_+}{(r_+^2+a^2)^2},\qquad \bar{k}_{\psi} = \frac{2b r_+}{(r_+^2+b^2)^2}.
\end{align}
Here
\correctedA{$r_+^2=ab$}
and $\rho_+^2 =r_+^2 +a^2\cos^2\theta+b^2\sin^2\theta$.
This corresponds to the near horizon geometry of \correctedA{the} extremal Myers-Perry black hole.
The two angular momenta $J_{\phi}$, $J_{\psi}$ corresponding to $\phi$, $\psi$ are 
\begin{align}
J_{\phi}  =\frac{\pi}{4}a(a+b)^2, \qquad J_{\psi} = \frac{\pi}{4}b(a+b)^2,
\end{align} 
respectively. 

\subsection{Slow rotating Kaluza-Klein black hole}
On the other hand the relation to the extremal slow rotating Kaluza-Klein black hole is 
as follows. Let us first define $p$, $q$ and $j$ such that 
\begin{align}
p = \frac{1}{c_0^2}\sqrt{c_1\left(1-\frac{c_2}{c_0^2}\right)}, 
\qquad q^2 = \frac{c_1}{c_2{}^2}\left(1-\frac{c_2}{c_0^2}\right), \qquad
j^2 = 1+\frac{4c_0^2c_2}{c_1 ^2}.
\end{align}
By changing coordinates as 
\begin{align}
\cos\theta &= \frac{2\sigma-\sigma_1-\sigma_2}{\sigma_2-\sigma_1},\\
x^1 &= \frac{\sqrt{1-j^2}}{c_0^2j}\phi, 
\quad x^2 = -\frac{2}{c_0^2q}\sqrt{\frac{(p+q)}{p(1-j^2)}}
\left(
\frac{\phi}{\eta}-\correctedA{\sqrt{\frac{p+q}{p^3}}y}
\right),
\end{align} 
the horizon data is written as  
\begin{align}
  \gamma_{ij}dx^idx^j &= \frac{H_q}{H_p}(dy-A_{\phi}d\phi)^2 +\frac{(pq)^3(1-j^2)\sin^2 d\phi^2}{4(p+q)^2H_{q}}, 
\\ 
  \frac{\sigma}{Q(\sigma)}d\sigma^2 &= H_{p}d\theta^2,
  \qquad
  \Gamma = \frac{2(p+q)}{(pq)^{3/2}(1-j^2)^{1/2}}H_p, \\ 
  \bar{k}_{\phi} &= \frac{2(p+q)}{\,(pq)^{3/2}(1-j^2)},
  \qquad \bar{k}_y = \frac{2}{1-j^2}\sqrt{\frac{p+q}{q^3}},
\end{align}
with
\begin{align}
H_{p} &= \frac{p^2q}{2(p+q)}(1+j\cos\theta),\quad 
H_q = \frac{pq^2}{2(p+q)}(1-j\cos\theta),\\
A_{\phi} &= \frac{q^2p^{5/2}}{2(p+q)^{3/2}H_{q}}(j-\cos\theta).
\end{align}
When periodicities $\phi\sim\phi+2\pi$, $y\sim y+8\pi\tilde{P}$ are imposed,
this corresponds to the near horizon geometry of \correctedA{the} extremal slow rotating Kaluza-Klein black hole.
Then magnetic charge $\tilde{P}$, electric charge $\tilde{Q}$ and angular momentum $J$ are written as  
\begin{align}
\tilde{P}^2 = \frac{p^3}{4(p+q)}, \quad \tilde{Q}^2 = \frac{q^3}{4(p+q)}, 
\quad G_{4}J= \frac{(pq)^{3/2}}{4(p+q)}j.
\end{align}



\providecommand{\href}[2]{#2}\begingroup\raggedright\endgroup

\end{document}